\documentclass[sigconf]{acmart}

\AtBeginDocument{%
  }
\setcopyright{acmlicensed}
\copyrightyear{2025}
\acmYear{2025}
\setcopyright{rightsretained}
\acmConference[UIST Adjunct '25]{The 38th Annual ACM Symposium on User Interface Software and Technology}{September 28-October 1, 2025}{Busan, Republic of Korea}
\acmBooktitle{The 38th Annual ACM Symposium on User Interface Software and Technology (UIST Adjunct '25), September 28-October 1, 2025, Busan, Republic of Korea}\acmDOI{10.1145/3746058.3758422}
\acmISBN{979-8-4007-2036-9/2025/09}

\begin{document}

\title{ConflictLens: LLM-Based Conflict Resolution Training in Romantic Relationship}

\author{Jiwon Chun}
\authornote{Both authors contributed equally to this research.}
\email{jiwonchun@tamu.edu}
\orcid{0009-0002-1516-7586}
\affiliation{%
	\institution{Texas A\&M University}
	\city{College Station}
	\state{Texas}
	\country{USA}
}

\author{Gefei Zhang}
\authornotemark[1]
\email{gefei@zjut.edu.cn}
\orcid{0009-0005-5704-9176}
\affiliation{%
	\institution{Zhejiang University of Technology}
	\city{Hangzhou}
	\state{Zhejiang}
	\country{China}
}

\author{Meng Xia}
\email{mengxia@tamu.edu}
\orcid{0000-0002-2676-9032}
\affiliation{%
	\institution{Texas A\&M University}
	\city{College Station}
	\state{Texas}
	\country{USA}
}

\renewcommand{\shortauthors}{Jiwon Chun, Gefei Zhang, and Meng Xia}

\begin{abstract}

Our poster presents \textit{ConflictLens}, a three-stage simulation system powered by large language models (LLMs) and grounded in psychological theory, designed to help users reflect on and practice conflict resolution in romantic relationships. Users can upload real conflict scenarios to receive evaluation of behavioral patterns, reflect on conflicts by annotating their negative behaviors, and practice different conflict resolution strategies in AI-simulated duologues.
Initial evaluation by three domain experts suggests that \textit{ConflictLens} offers a realistic experience and effectively supports self-guided reflection and communication practice in romantic relationships.

\end{abstract}



\maketitle
\section{Introduction and Related Work}
Romantic conflict is deeply emotional, shaped more by personal histories and relational dynamics than by rational decision-making. Unlike everyday disagreements, these conflicts often stem from complex factors such as dependency, power imbalances, emotional manipulation, and unmet expectations~\cite{suler2011psychology, meyer2022relationship}. Prior work highlights this emotional complexity, for example, Lily embeds romantic lyrics into couples' conversations to enhance emotional expression~\cite{kim2019love}. However, responses to conflict such as withdrawal, anger, or over-accommodation are often shaped by behavior patterns, which are rooted in attachment, communication, and coping styles. If not addressed, these patterns can damage trust, hinder communication, and escalate into emotional abuse~\cite{huang2024no}.

Recent work has explored the use of large language models (LLMs) to support romantic communication. For instance, Rehearsal \cite{shaikh2024rehearsal} uses LLMs to generate dialogue scripts that help users practice communication techniques through scripted ``rehearsals.'' However, such tools emphasize surface-level skills, without addressing the deeper behavioral patterns and emotional dynamics that drive conflict. Baughan et al.~\cite{baughan2024supporting} proposed a temporal model for how digital interventions might support users in difficult conversations, but it has yet to be realized in a working system. While these studies show promise in improving communication~\cite{louie2024roleplay}, they largely overlook users' need to understand why conflicts arise in the first place. This leads to our research question: How can we help users understand the behavioral and emotional roots of romantic conflict and guide them toward reflection through interactive practice?

\begin{figure}
    \centering
    \includegraphics[width=1\linewidth]{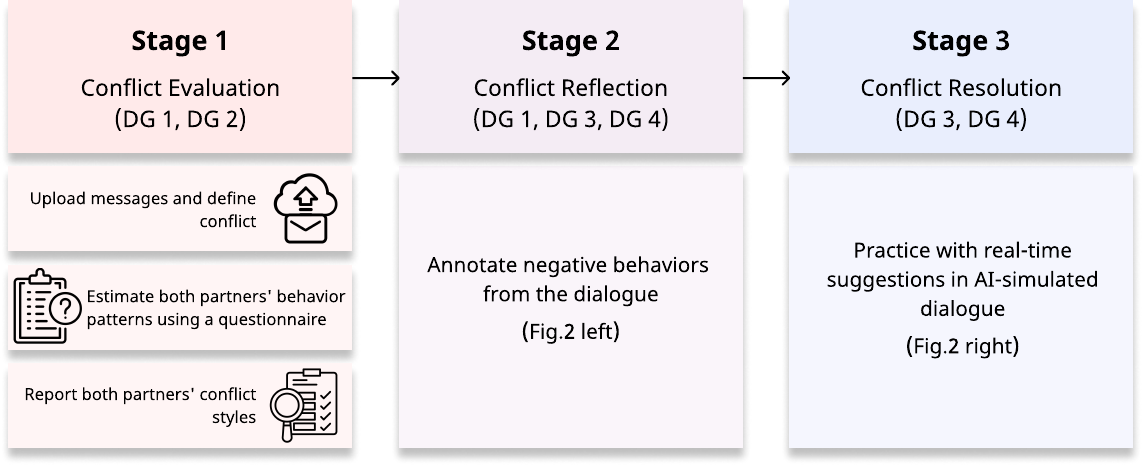}
    \caption{Three-stage flow of \textit{ConflictLens}. Stage 1 (DG 1, 2): Upload messages, estimate questionnaire responses, and classify conflict styles to evaluate the conflict. Stage 2 (DG 1, 3, 4): Annotate dialogue to reflect conflict. Stage 3 (DG 3, 4): Practice with AI simulation to resolve conflict.}
    \Description{Three-stage flow of \textit{ConflictLens}. Stage 1 (DG1, 2): Upload messages, estimate questionnaire responses, and classify conflict styles to evaluate the conflict. Stage 2 (DG1, 3, 4): Annotate dialogue to reflect conflict. Stage 3 (DG 3, 4): Practice with AI simulation to resolve conflict.}
    \label{fig:system}
    \vspace{-2em}
\end{figure}

\begin{figure*}[t]
\includegraphics[width=\textwidth]{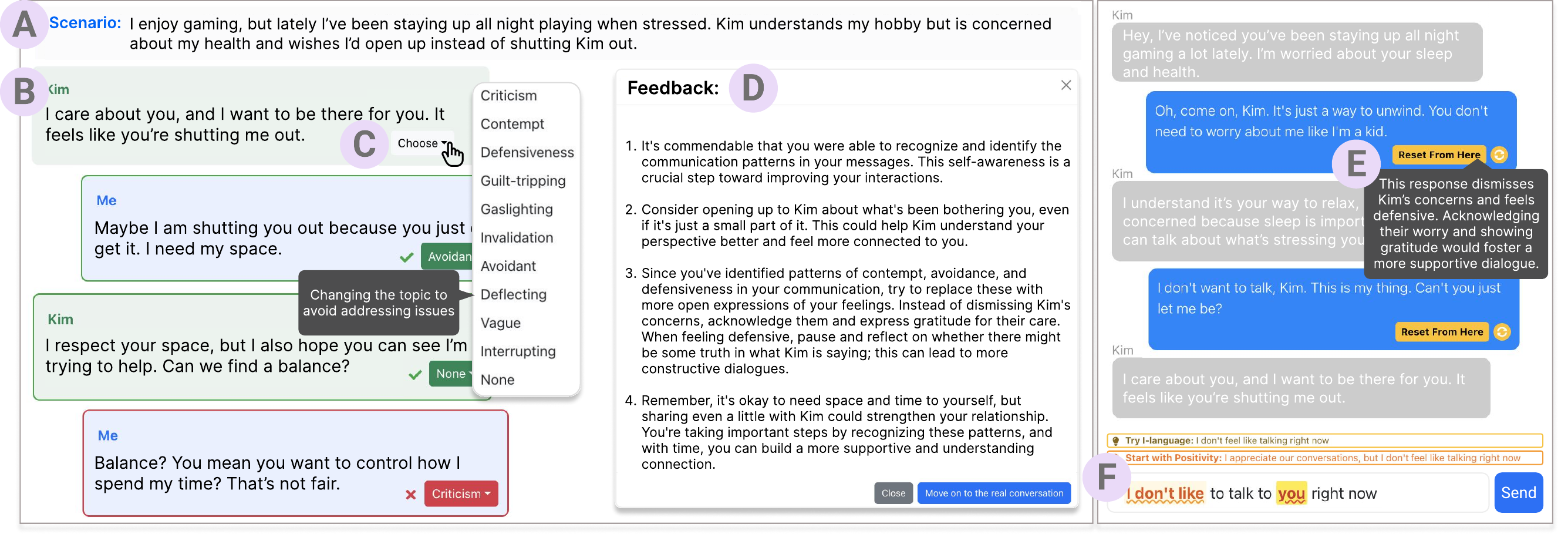}
  \caption{\textit{ConflictLens} interface: Stage 2 (left) and Stage 3 (right). (A) Realistic conflict scenario. (B) Simulated conversations between the user and the partner. (C) Dropdown to annotate negative communication behaviors (D) Feedback on annotations. (E) Recommended reset point with ``Reset From Here'' button. (F) Real-time rewrite suggestions.}
  \Description{\textit{ConflictLens} interface: Stage 2 (left) and Stage 3 (right). (A) Realistic conflict scenario. (B) Simulated conversations between the user and the partner. (C) Dropdown to annotate negative communication behaviors (D) Feedback on annotations. (E) Recommended reset point with ``Reset From Here'' button. (F) Real-time rewrite suggestions}
  \label{fig:teaser}
\end{figure*}
\section{Formative Study}
We conducted formative interviews with three experts in the research of interpersonal relationships (E1-E3; mean experience: 8.33 years, $SD = 9.29$) to inform the design of our system and deepen our understanding of how people perceive, experience, and resolve romantic conflict. We conducted 30-minute Zoom interviews, approved by the IRB, and provided each participant with a \$10 Amazon gift card. Interview questions are included in the supplementary material.
From the interview results, we derive four design goals.

\noindent\textbf{DG 1.} Support users in identifying the root causes of conflict by understanding both partners’ needs and communication styles (E1–E3).

\noindent\textbf{DG 2.} Encourage users to reflect on past conversations to gain deeper insights (E1–E3).

\noindent\textbf{DG 3.} Provide personalized AI-driven simulations to help users practice conflict responses and improve future interactions~\cite{kim2024opportunities} (E3).

\noindent\textbf{DG 4.} Offer actionable feedback that fosters ongoing reflection and learning (E1, E3).
\section{System Design and Implementation}
Based on our design goals and psychological theory ~\cite{fisher2011getting}, \textit{ConflictLens} guides users through three stages (Fig.\ref{fig:system}), starting with mutual understanding, then moving to reflection and practice. All the features below are driven by the OpenAI API with prompt engineering.

\textbf{Stage 1: Conflict Evaluation (DG 1, DG 2)}
Users start by uploading screenshots of conflict texts. \textit{ConflictLens} then estimates the behavior patterns of both partners based on a 13-item conflict resolution questionnaire~\cite{zacchilli2009romantic} (5-point Likert scale), which users can review and adjust. Based on the finalized scores, \textit{ConflictLens} classifies both conflict styles (Avoidant, Validating, Volatile, or Hostile~\cite{busby2009perceived, zheng2021pocketbot}), highlights negative communication patterns.

\textbf{Stage 2: Conflict Reflection (DG 1, DG 3, DG 4)}
Based on the partners' identified styles from Stage 1 (e.g., Hostile vs. Validating), \textit{ConflictLens} generates a 15-turn AI-simulated dialogue on common topics such as household habits~\cite{meyer2022relationship} (Fig.\ref{fig:teaser} A). Users label each utterance with one of the 11 negative communication behaviors~\cite{gottman2012marriages,garcia2023we,petronio2010communication,brown2014politeness} (Fig.\ref{fig:teaser} C) and receive instant feedback on their accuracy. After annotation, the system provides a concise summary of their strengths in recognizing their negative behavior patterns and tailored recommendations for improvement (Fig.\ref{fig:teaser} D).

\textbf{Stage 3: Conflict Resolution (DG 3, DG 4)}
Within the ongoing scenario from Stage 2, users can continue the conversation or restart the conversation from the recommended points (Fig.\ref{fig:teaser} E). The system will simulate the partner's responses based on the styles identified in Stage 1. As users try different ways to communicate, \textit{ConflictLens} provides real-time suggestions (Fig.\ref{fig:teaser} F) to promote non-violent communication such as using I-language and starting with positive framing~\cite{baughan2024supporting,caraban201923}.
\section{Evaluation}
The same three experts (E1-E3) from the formative study evaluated the deployed system in a 30-minute Zoom session, each receiving a \$10 Amazon gift card. 
They were given a conflict scene screenshot from the reality TV show \textit{See you again}, uploaded it to the system, explored its features, and shared their feedback. Screenshots and questionnaire are in the supplementary material.

\textbf{Usability} Participants found the interface clear and easy to follow. E2 noted that it was easy to become immersed in the system's logic and explore the root causes of the conflict. He especially valued the during- and post-conflict analysis for clarifying the situation.
Although E1 and E2 found some pages text-heavy, they mentioned interactive features such as Fig.\ref{fig:teaser} C and Fig.\ref{fig:teaser} E made the content more manageable and engaging.

\textbf{Effectiveness} Participants felt the workflow was generally effective. E2 valued the conflict type analysis results in Stage 1 because it helps users to understand how to approach conflict. E1 and E3 found the simulated conversation in Stage 3 (Fig.\ref{fig:teaser} right) helpful, as it allowed them to try different branches of dialogue. E1 highlighted the real-time suggestions Fig.\ref{fig:teaser} F as useful. 

\textbf{Simulation Quality}
Participants found the AI-simulated dialogues realistic. E1 observed that the screenshot depicted a habit-related scenario, which was reflected in the AI-generated dialogue. E3 also found the simulated conversation good and contextually appropriate, although it was not based on their own screenshots.
\section{Discussion and Future Work}
\textit{ConflictLens} shows promise, but several limitations remain. First, our formative study involved only three domain experts, limiting generalizability. Larger-scale studies with diverse users are needed to evaluate usability and emotional relevance. Second, current evaluations rely on users' self-assessments and do not confirm whether the simulated dialogues reflect real-life communication. Future work could include partner feedback to assess realism. Third, uploading private conversations raises privacy and ethical concerns~\cite{kim2019love,kim2024opportunities}. Future deployments should ensure mutual consent and apply safeguards such as on-device redaction, guided by value-sensitive design~\cite{shelby2023sociotechnical,showkat2022s,friedman1996value}. Finally, overreliance on AI in interpersonal contexts may lead to emotional dependency. We plan to add usage safeguards and conduct longitudinal studies to assess long-term impact and edge cases.

\bibliographystyle{ACM-Reference-Format}
\bibliography{References}
\end{document}